\documentclass[10pt,letterpaper]{article} 
\usepackage{opex3}
\usepackage[latin1]{inputenc}
\usepackage[reqno]{amsmath}
\usepackage{amssymb}

\newcommand{\un}[1]{\mathrm{\;#1}}
\DeclareInputMath{181}{\mu}
\DeclareInputMath{183}{\cdot}

\begin{document}

\title{Transmittance and near-field characterization
of sub-wavelength tapered optical fibers}

\author{Fedja Orucevic, Valérie Lefèvre-Seguin, Jean Hare$^*$}

\address{École Normale Supérieure; Univ. Pierre et Marie Curie--Paris 6;  CNRS\\
Laboratoire Kastler Brossel -- 24 rue Lhomond -- 75231 Paris cedex 05 -- France
\vskip-2mm
\email{$^*$Jean.Hare@lkb.ens.fr}}
\vskip-4mm
\begin{abstract}
We have produced high transmission sub-wavelength tapered optical fibers for the purpose of whispering gallery mode coupling in fused silica microcavities at 780~nm. A detailed analysis of the fiber transmittance evolution during tapering is demonstrated to reflect precisely the mode coupling and cutoff in the fiber. This allows to control the final size, the number of guided modes and their effective index. These results are checked by evanescent wave mapping measurements on the resulting taper.
\end{abstract}
\ocis{(060.0060) Fiber optics; (060.2270) Fiber characterization; (140.3948) Microcavity devices; (140.4780) Optical resonators}


\section*{Introduction}
Tapered optical fibers have attracted a large interest in the two last decades, due to their numerous applications as directional fiber-fiber couplers, filters for wavelength multiplexing\cite{GonthierLacroix89}, biosensors and more recently for the production of broad-spectrum-light from femtosecond laser impulsions\cite{BirksWadsworth00}. They have also been used successfully to selectively excite the  whispering gallery modes (WGM) of spheroidal or toroidal microcavities \cite{KnightCheungt97,SpillaneKippenberg03}.  The need for low losses has often led to work with waist diameters in the range $5$ to $50\un{µm}$ while the weakness of the evanescent wave was, if needed, compensated by long interaction lengths. Efficient coupling to microcavities WGMs implies interaction lengths reduced to a few micrometers which in turn requires to work with smaller taper diameters and to control the effective index of the taper modes. The mode coupling and the resulting energy transfer have been extensively analyzed in the literature \cite{Snyder70,LoveHenry91,MoarHuntington99}, either by Beam Propagation Method or, for slowly varying tapers radii, applying the coupled mode theory. The latter approach is best suited for tapers with characteristic lengths  long enough to provide a physical meaning to the so called ``local modes", defined at each position as the modes guided by a cylindrical fiber with the same diameter.

In this letter we describe how we produced sub-wavelength nearly adiabatic tapers with an overall transmittance higher than 90\%, thanks to very efficient adiabatic transfer from the single mode of the untapered fiber to the fundamental mode of the central part of the taper.  In the frame of ``local modes'' modelling, we show that the final properties of the taper can be precisely deduced from the transmittance evolution during the tapering process. Moreover an experimental criterion for a single-mode taper is demonstrated.
Near-field mapping measurements of the tapers obtained by this method confirm our model.

\section*{Taper fabrication}
\begin{figure}[hbt]
 \centering
 \includegraphics[width= 10cm]{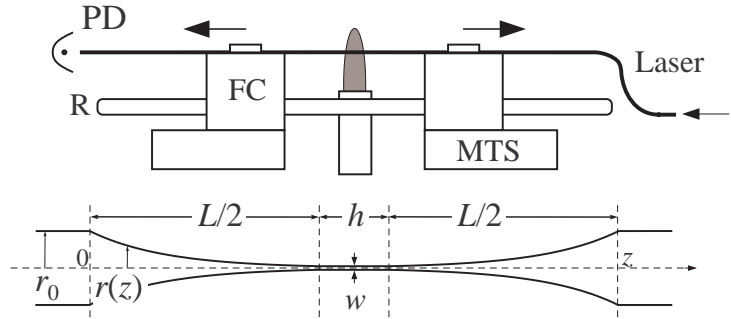}
 \caption{Top: Sketch of the experiment, showing the microtorch in the center, the two fiberclamps FC, the rod R, the motorized translation stages MTS, the photodiode PD. Bottom: Shape of the resulting taper (not to scale), with definition of  abscissa $z$, lengthening $L$, hot-zone $h$, taper waist $w$, initial radius $r_0$.}
 \label{f:setup}
\end{figure}

Our experimental setup is sketched in Fig.~\ref{f:setup}. The single mode FS-SN4224 fiber is held by two fiberclamps, sliding along two parallel stainless steel rods. The fiber is heated to its softening point by a butane microtorch, using a specially designed nozzle producing a short flame which is $\approx 10\un{mm}$ wide along  the fiber axis. The fiber position  in the flame is a critical parameter which can be adjusted by moving the nozzle using a three axis translation stage. The two fiberclamps are then symmetrically moved apart at the same velocity $v$ of about $40 \un{\mu m·s^{-1}}$ by two motorized translation stages. During the whole process, the fiber transmission is monitored using a laser at the working wavelength $\lambda\approx775\un{nm}$ and a photodiode (PD).
\begin{figure}[hbt]
 \centering
 \includegraphics[width= 10cm]{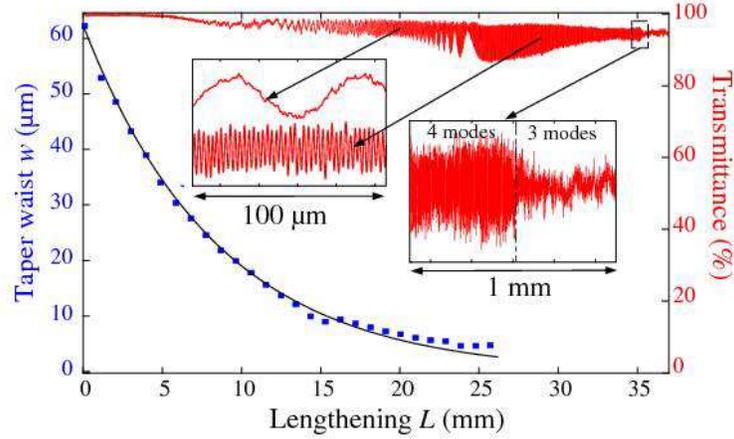}
 \caption{Fiber transmittance as a function of the fiber lengthening $L$ (right scale). Note the 95\%  final transmission (ie $-0.22\un{dB}$ insertion loss).
The curve in solid squares is the observed decrease of the fiber waist $w$ (left scale). Left inset: zoom on the transmittance curve over $100\un{µm}$ for $L=20\un{mm}$ and  $L=30\un{mm}$. Right inset: zoom on the last amplitude drop.}
 \label{f:observ}
\end{figure}

Figure \ref{f:observ} shows a typical recording of the fiber transmission during the tapering process, which lasts about 6 minutes, as a function of the fiber lengthening $L=2vt$. After a slight decrease of the transmission, oscillations appear, the  frequency of which increases with time, as shown in the left inset. These oscillations are known to result from the interference between different modes and the modulation of their envelope  is due to the beating of different frequencies. The most important feature is the amplitude drop occurring (for this taper) at $L=35\un{mm}$.  This is the experimental signature that the taper becomes nearly single mode, as shown below.

We have also observed with a video-microscope the exponential decrease of the taper waist, shown in solid squares in Fig~\ref{f:observ}. Note that the experimental points stops before the end of the tapering; indeed, a transverse motion of the thinned fiber in the turbulent microtorch gas flow, made the last image too fuzzy for a meaningful fiber diameter measurement. The experimental curve is well fitted by a exponential, as predicted by the model given in \cite{BirksLi92}, relying on volume conservation. According to this model, the taper shape in cylindrical coordinates, with origin $z=0$ at one end of the taper (see Fig.~\ref{f:setup}), is described by:
\begin{equation}\label{e:radius}
  r(z,L)=\left\{\begin{aligned}
        &r_0\, \exp(-z/h) \quad &\text{for}\ &&0 &<z<\tfrac{L}{2}  \\
        &w\equiv r_0\, \exp(-L/2h)\quad &\text{for}\ &&\tfrac{L}{2}&<z<\tfrac{L+h}{2}
     \end{aligned} \right.
\end{equation}
with a symmetric profile for $z>(L+h)/2$. In this equation $r_0=62.5\un{µm}$ is the initial fiber radius, $L$ is the fiber lengthening and $h$ the so called ``hot zone'' length. The ``hot zone'' is the cylindrical part where the softened fiber is drawn, roughly corresponding to the flame width. Though sensitive to the fiber location in the flame, the fitted $h$ for adiabatic tapers remains in the range 6--$8\un{mm}$, consistent with our flame geometry.

\section*{Analysis of oscillations}

The observed oscillations originate from the beating of different local modes, which propagate along the taper with different propagation constants $\beta_i\equiv 2\pi/\lambda×N_\text{eff,\,i}$ ($N_\text{eff,\,i}$ is their effective index), and recombine at the output with a relative phase which depends on the effective length of the taper. The amplitude of these oscillations varies during the lengthening of the fiber and reveals the efficiency of excitation and recombination of the different modes during the tapering process. In Fig~\ref{f:observ}, a peak-to-peak amplitude of 10\% corresponds to an energy transfer of only 5\% from the fundamental to excited modes, resulting in a final transmission of 95\%.

\begin{figure}[hbt]
 \centering
 \includegraphics[width= 10cm]{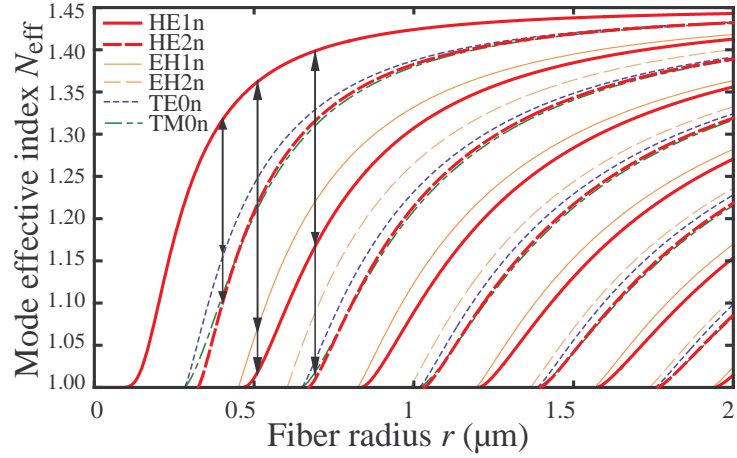}
 \caption{Calculated effective index for the lowest modes of a cylindrical silica fiber, as a function of fiber radius. The thick solid and dashed lines are for $\mathrm{HE}_{1n}$  and $\mathrm{HE}_{2n}$ modes; the thin solid and dashed lines for the $\mathrm{EH}_{1n}$ and $\mathrm{EH}_{2n}$ modes, the thin dotted lines for the $\mathrm{TE}_{0n}$ modes.}
 \label{f:Neff}
\end{figure}

Figure~\ref{f:Neff} shows the effective indices of the first modes of a cylindrical waveguide with a uniform silica core and an air or vacuum cladding ($N=1$). Due to strong guiding conditions, the relevant modes differ from the familiar LP modes, and include ``hybrid modes" with more complex polarization properties. In particular, the fundamental mode is the hybrid $\mathrm{HE}_{11}$ mode \cite{SnyderLove}. For large radii the modes effective indices are close to the maximal value $N_S=1.453$ owing to a good confinement in silica and small numeric aperture. On the opposite for small radii the air penetration increases thus diminishing the effective index down to the minimal value $\min(N_\text{eff})=1$, which is reached at mode cutoff. As a result, the effective index dispersion increases as the taper radius decreases, hence the gradual increase of the oscillation frequency.

Further interpretation of the oscillations is obtained by analyzing the beating frequency of two given local modes, designated by indices $i=1$ or 2. Note that due to the taper radius variation along the propagation axis $z$, the associated propagation constants $\beta_i(r)$ are $z$--dependent. Using the symmetry around the taper center, the accumulated relative phase writes:
\begin{equation} \label{e:phase}
 \Phi_{12}(L) = 2\left\{\ \int_0^\frac{L}{2} \Delta\beta_{12}(r(z))\;dz \ + \  \Delta\beta_{12}(w)\frac{h}{2}  \right\} \ .
\end{equation}
 A simple derivation taking into account the exponential variation of $w$ vs. $L$, leads to the spatial angular frequency :
\begin{equation} \label{e:freq}
  K_{12}=\frac{d\Phi_{12}}{dL} = \Delta\beta_{12}(w)-\frac{w}{2}\left.\frac{d}{dr}\left( \Delta\beta_{12}\right)\right|_{w} \ .
\end{equation}

\begin{figure}[hbt]
 \centering
 \includegraphics[width=10cm]{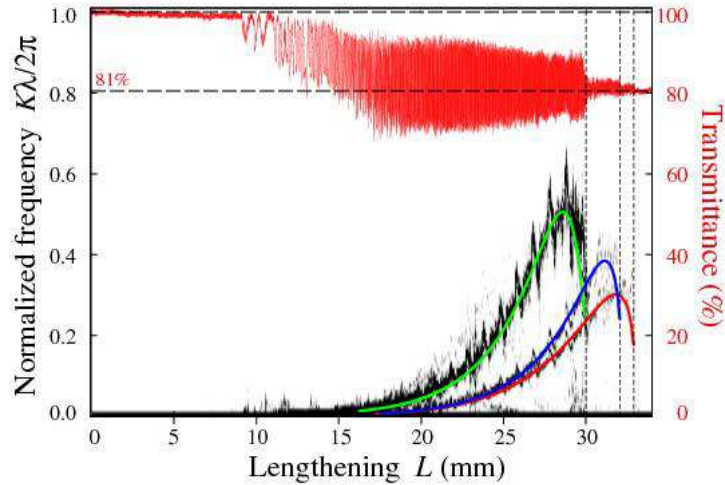}
 \caption{Transmittance curve and its short-time Fourier transform (window width is set to $\sim 0.5\un{mm}$). The solid curves are the frequencies calculated from eq.~(\ref{e:freq}) with $h$ as single fit parameter. The vertical lines underline the coincidence of amplitude drop with mode cutoff.}
 \label{f:sonogram}
\end{figure}

This frequency can be derived by performing a short-time Fourier transform of the beating, known in acoustics as a sonogram or spectrogram \cite{wiki_spectrogram}. Figure~\ref{f:sonogram} shows such a spectrogram with the lengthening as abscissa and the normalized frequency $K\lambda/2\pi$ as ordinate. It should be emphasized that a poorly adiabatic taper (with $h<5\un{mm}$) has been intentionally chosen in order to make the discussed features more visible. In this figure is also plotted the normalized frequency calculated from eq.~(\ref{e:freq}) and eq.~(\ref{e:radius}). They correspond to the beating of the fundamental mode $\mathrm{HE}_{11}$ with the $\mathrm{HE}_{12}$, $\mathrm{HE}_{21}$ and $\mathrm{TE}_{01}$ modes, from the highest to the lowest curve, respectively. Using the ``hot zone'' width $h$ as an adjustable parameter, our model reproduces accurately the experimental observations, with the fitted value $h_0=3.05\un{mm}$.

As expected the most intense component arises from the beating of the fundamental mode with the $\mathrm{HE}_{12}$ mode, the first excited mode that has the same symmetry (which is also the symmetry of the LP01 mode initially launched in the single mode fiber). The corresponding difference $\Delta N_\text{eff}$ is illustrated on Fig~\ref{f:Neff} by the righter-most arrow.
Note that the small decrease of the spatial frequency just before cutoff arises from the inflection point of the dispersion curves visible in Fig.~\ref{f:Neff}, and of the presence of the derivative term in eq.~(\ref{e:freq}).

After the cutoff of the $\mathrm{HE}_{12}$ mode (occurring at $w\approx 460\un{nm}$) the amplitude of the remaining oscillations is reduced to a few \%, corresponding to the two lighter lines on the sonogram. They result from the weak excitation of two modes with a different symmetry, which can likely be attributed to imperfections in the early stage of tapering.  When these two modes successively reach cutoff, the oscillations fully disappear. From then on the taper is singlemode, with a waist smaller than 300~nm.

\begin{figure}[hbt]
 \centering
 \includegraphics[width=100mm]{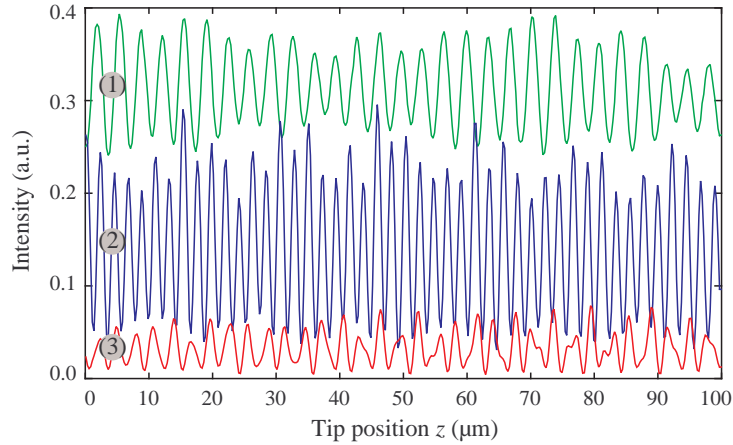}
 \caption{Near-field intensity mappings recorded (1) at the taper center , 
 (2) 1.25~mm before the hot zone  and (3) 2.25~mm before the hot zone .}
 \label{f:mapping}
\end{figure}

In order to confirm our interpretation of the oscillation pattern, we mapped the evanescent field of the taper in its final shape, similarly to the mapping experiment of the whispering gallery mode field reported in \cite{KnightDubreuil96}. A thin fiber tip is scanned along the taper using a 3-axis piezo-translation stage while the intensity catched by the fiber tip is monitored. Looking only to the intensity distribution along the taper, we put the tip in contact with the taper. This ensures a good mechanical stability and allows to scan over the whole $100\un{µm}$-stroke of the piezo stage.

Figure \ref{f:mapping} shows the detected signal as a function of position for three different places along the taper studied in Fig.~\ref{f:observ}: (1) in the central region, (2) at a distance 1.25~mm before the hot zone and (3) 2.25~mm before the hot zone. As expected, the average intensity becomes smaller when the taper becomes thicker, because the modes are more efficiently confined inside the silica. We also observe that the fringes visibility is much larger than on the taper transmittance curve: indeed the excited mode that carries only a few percent of the total power is less confined than the fundamental mode, and has therefore a larger relative intensity in its evanescent part. We performed a Fourier transform of these signals in order to identify the involved modes. The results are given in table~\ref{t:fourier}, retaining in the detected signal and for each scan the two principal modes beating with the fundamental. The weights in table~\ref{t:fourier} give their relative contribution in intensity to the fiber tip signal. The periods $\Lambda$ are recast in effective index differences by the relation $\Delta N_{eff}=\lambda/\Lambda$, and the taper radius is determined by the co-existence of the two effective index differences taken into account (see black arrows in Fig.~\ref{f:Neff}).

\begin{table}\centering
\caption{Frequency analysis of Fig.~\ref{f:mapping}} \label{t:fourier}
\begin{tabular}{lllllll}
\hline
\strut & \multicolumn{2}{c}{Scan 1} & \multicolumn{2}{c}{ Scan 2} & \multicolumn{2}{c}{Scan 3} \\
\hline
Offset position (mm)& \multicolumn{2}{c}{0} & \multicolumn{2}{c}{ 1.25 mm} & \multicolumn{2}{c}{2.25 mm} \\
Weight & 95\% & 4\% & 92\% & 3\% & 72\% & 20\% \\
$\Delta N_\text{eff}$ & 0.22 & 0.16 & 0.35 & 0.30 & 0.26 & 0.40 \\
Beating mode & HE21 & TE01 & HE12 & EH11 & HE12 & HE22 \\
Radius $r (\un{µm})$ & \multicolumn{2}{c}{$0.4$}  & \multicolumn{2}{c}{$0.51$}& \multicolumn{2}{c}{0.68}   \\
\hline
\end{tabular}

\end{table}

\section*{Conclusion}

We have shown that the oscillations of the transmitted power during the tapering  process provides a real-time information on the modes propagating in the cylindrical part of the taper. The beating frequency is directly related to the associated effective indices. The successive drops observed on their intensity in the final stage is a convenient criterion to ascertain the cutoff of the last guided modes, and provides a mean to directly control the final taper diameter, in a size range where a direct optical measurement of the fiber diameter is no longer possible. Moreover this allows to control the fundamental mode effective index between 1.4 and 1.25, which is the appropriate range for microsphere and microtoroid's whispering gallery modes excitation.

\section*{Acknowledgements} The authors acknowledge fruitful discussions with Gilles Nogues and Jean-Michel Gérard.

\end{document}